# Tuning the Electronic States of $Bi_2Se_3$ Films with Large Spin-Orbit Interaction Using Molecular Heterojunctions

M. Rogers,[1] C. Knox,[1] B. Hickey,[1] L. Ansari,[2] F. Gity,[2] T. Moorsom,[1,3] M. McCauley,[4] G. Teobaldi,[5] M. dos Santos Dias,[6] H. B. Vasili,[1] M. Valvidares,[7] M. Ali,[1] G. Burnell,[1] A. Yagmur,[1] S. Sasaki[1] and O. Cespedes[1,†]

1. School of Physics and Astronomy, University of Leeds, Leeds, U.K.
2. Micronano Electronics Group, Tyndall National Institute, University College Cork, Cork, Republic of Ireland
3. School of Chemical and Process Engineering, University of Leeds, Leeds, U.K.
4. SUPA, School of Physics and Astronomy, University of Glasgow, Glasgow, U.K.
5. Scientific Computing Department, Science & Technology Facilities Council UKRI, Rutherford Appleton Laboratory, Didcot OX11 0QX, U.K.
6. Scientific Computing Department, Science & Technology Facilities Council UKRI, Daresbury Laboratory, Warrington WA4 4FS, United Kingdom
7. ALBA Synchrotron Light Source, 08290 Cerdanyola del Vallès, Barcelona, Catalonia, Spain
[†]Corresponding author email address: o.cespedes@leeds.ac.uk

**An electric bias can shift the Fermi level along the Dirac cone of a topological insulator and modify its charge transport, but tuning the electronic states and spin-orbit interaction (SOI) without destroying the surface topology is challenging. Here, we show that thin film $Bi_2Se_3$/n-p (p-n) molecular diodes form ordered interfaces where charge transfer and orbital re-hybridisation result in a decrease (increase) of the carrier density and improved mobility. In $Bi_2Se_3$ the spin-orbit lifetime, $\tau_{so}$, is 0.13 ps, which is comparable to the strongest spin-orbit materials. This lifetime drops further to 0.06 ps (0.09 ps) with the addition of p-n (n-p) molecular diodes, at the limit of measurable values. This strengthened spin-orbit interaction occurs even though molecules are made of light elements and increase the mean free path of the charge carriers by almost 50%, indicating changes to the Berry curvature and/or Rashba splitting around the hybridisation points. Raman spectroscopy gives evidence that the coupling effect may be controlled by optical irradiation, opening a pathway towards the design of heavy-light element hybrids with optically tunable quantum transport.**

I. Introduction

The tunability of charge transport in Dirac materials such as the Topological Insulator (TI) $Bi_2Se_3$ is one of their most attractive aspects for applications in conventional electronics. With a dielectric gate it is possible to control the carrier type, concentration and mobility. However, processes to add oxide electrodes may result in damage of the surface states. Furthermore, the influence of the electrical bias is restricted to shifting the Fermi level along the Dirac cone without altering the density of states, so biasing has a limited influence in the quantum transport properties of the material. TIs are materials where, ideally, the bulk of the sample is insulating and the surface states are conducting, topologically protected. The SOI is key to their magneto-electronic properties, such spin-momentum locking and spin dependent electron scattering that can be converted into an electric voltage via spin accumulation and the imbalance in chemical potential for each orientation -the inverse spin Hall effect. The SOI then controls applications such as the photo-galvanic effect, spin transfer torque in magneto-resistive random-access memories (STT-MRAM), [1–4] unidirectional spin Hall effect, [5] weak anti-localisation and polarised dynamics. Some of the issues that prevent the adoption of Dirac materials in technologies include interface and electrical contact problems, with resistivity mismatch resulting in charge accumulation and scattering, interdiffusion and lack of SOI tunability. [6–8]

Molecular materials offer an alternative to conventional bias gating with oxide electrodes by forming hybrid states that allow for spin filtering, spin photovoltaics and efficient spin-voltage conversion. [9] For example, sub-ML coverage of $Bi_2Se_3$ with Manganese-Phthalocyanine (MnPc) shifts the Dirac point while the topological states remain unaffected. [10] FePc grown on $Bi_2Te_3$ gives rise to a supermolecular lattice where the magnetic moment of Fe(II) persists with an in-plane magnetic easy axis without destroying the topological surface states. [11] $C_{60}$ overlayers result in the formation of dispersive conduction bands. [12] Growth of F4TCNQ and Co(acac)3 on $Bi_2Se_3$ results in a higher stability in atmosphere, with a mobility enhanced and sustained for months in ambient conditions. [13] Furthermore, organic materials, despite being made mostly of carbon and hydrogen, can give rise to strong spin-orbit effects, either due to their intrinsic curvature, via interface Rashba effects, by the inclusion of metal centres, or due to charge transfer and hybridisation. [14–17] In order to enhance and tune these effects, we use molecular bilayers where one of the materials is an n-type molecule with electron carriers and strong electron affinity ($C_{60}$) and the other is a p-type hole-conductor molecule (Mn phthalocyanine). The use of both molecules in a diode bilayer leads to a built-in bias that, combined with the charge transfer and hybridisation effects at the TI surface, can be used to tune not only the charge transport, but also the spin-orbit interaction and band structure of $Bi_2Se_3$.

Furthermore, because of the opto-electronic properties of the molecules, light irradiation can be used to tune these effects. The aim is to enhance spin-dependent electronic effects such as the Berry curvature, itself linked to symmetry-breaking interactions such as those present at the molecular interface. Altering the effective SOI and Berry curvature would enable the tuning of not only charge transport, but also weak antilocalisation, molecular dynamics, spin-torque and spin-voltage conversion effects amongst others.

## II. Structural and Interface Characterisation

We grew 14 nm thick $Bi_2Se_3$ films by molecular beam epitaxy on sapphire substrates, and molecular bilayers are then deposited in-situ without breaking vacuum by thermal evaporation -see Methods for details. The resulting multilayers are highly ordered, with both TI and molecular films showing crystalline order in cross-section transmission electron microscopy (TEM) images and Xray diffraction peaks (Figs. 1a-c). Atomic resolution, aberration corrected TEM images provide evidence for electron transfer at the $Bi_2Se_3$/$C_{60}$ interface, where local charged points can be observed as brighter molecular regions. The localised charges will act as local scattering points and, due to changes in orbital geometry and hybridisation, the Berry curvature may increase around these points. [18] Energy Filtered Transmission Electron Microscopy (EFTEM) elemental analysis of the structures shows little or no interdiffusion, although high resolution analysis of phthalocyanine layers is challenging due to beam damage even at low voltages/energies (Figs. 1d-e).

Supplementary Figure S1 shows the resistivity as a function of temperature, with the data extrapolated to zero Kelvin to determine $\rho_0$, the limit of elastic impurity scattering. The fits of the Bloch-Grüneissen function return the phonon resistivity and the Debye temperature. The overlayers did not change the phonon resistivity, but the Debye temperatures for the overlaid samples were a bit lower. In all samples the elastic mean free path ($l = v_F \tau_0$) is larger than the thickness and hence the samples are two-dimensional with respect to transport, with larger diffusion constants in the case of molecular diode-tuned states.

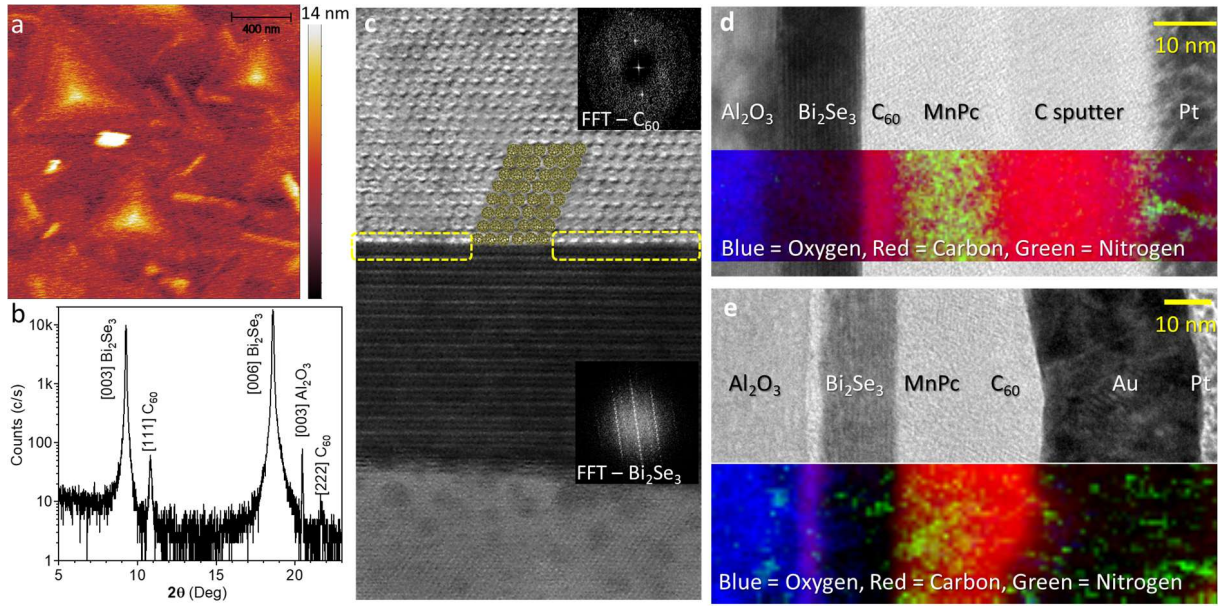

*Figure 1| Structural characterisation. a.* Cross-sectional Transmission Electron Microscope images of show a highly ordered $Bi_2Se_3/C_{60}$ interface with (111) FCC molecular crystal arrangement. *b.* Atomic Force Microscopy shows the triangular $Bi_2Se_3$ crystals on the surface. *c.* Xray Diffraction with the peaks corresponding to [001] $Bi_2Se_3$ and [111]. EFTEM sections for *d.* $Bi_2Se_3$/n-p and *e.* $Bi_2Se_3$/p-n show well-defined layers with little or no interdiffusion.

### III. Molecular Gating of Charge Transport

Similarly to an electric field bias, hybridisation and charge transfer at the molecular interface alter the carrier density and mobility of the system. Carrier density and mobility is determined via Hall effect measurements comparing devices cut from the same TI wafer but where only part is covered with a molecular bilayer. The organic diodes result in changes to the carrier density of $Bi_2Se$ that can be replicated by a top gate bias of ±10 V across a 24 nm thick $Al_2O_3$ dielectric layer. However, in the case of the alumina gate, the mobility and spin-orbit lifetime are not enhanced, while the carrier density increase is significantly less than for the molecular gate. Density Functional Theory calculations show changes in the band structure, including the possible emergence of flat bands that are precursor of strange metal behaviour. [19,20]

The acceptor-donor ($C_{60}$-MnPc) heterojunction generates a dipole dependent on the molecular arrangement. [21–23] Phthalocyanines grow edge-on when deposited on the fullerenes, which results in a charge transfer from $C_{60}$ to MnPc. We can model this heterojunction effect as a p-doped $C_{60}$ layer or with an in-built potential -Fig. 2e. A negative potential or p-doped molecule show a charge accumulation at the $Bi_2Se_3/C_{60}$ interface, in agreement with the increased interface brightness observed in TEM images. Whereas the carrier density of $Bi_2Se_3$ is reduced by 50% by a $C_{60}$ overlayer, the $C_{60}$/CuPc heterojunction reduces the carrier density just by 26% (Table 1). Fig. 2e illustrates the influence of an external electric field and molecular doping on charge transfer between the $C_{60}$ molecule and $Bi_2Se_3$. This graph represents charge redistribution and net charge transfer as functions of position along the z-direction, perpendicular to the surface. The result highlights how different external factors influence charge redistribution, where the peaks indicate charge accumulation, and the dips correspond to charge depletion. Similar to the charge density, total charge changes are very small far from the $C_{60}$/$Bi_2Se_3$ interface, indicating minimal charge movement in these regions (Fig. 2e bottom). The extent of charge transfer varies depending on the applied electric field and doping type. These variations reflect the combined effects of polarisation, dipole formation due to the electric field, and charge donation induced by p-type doping of the $C_{60}$ molecule.

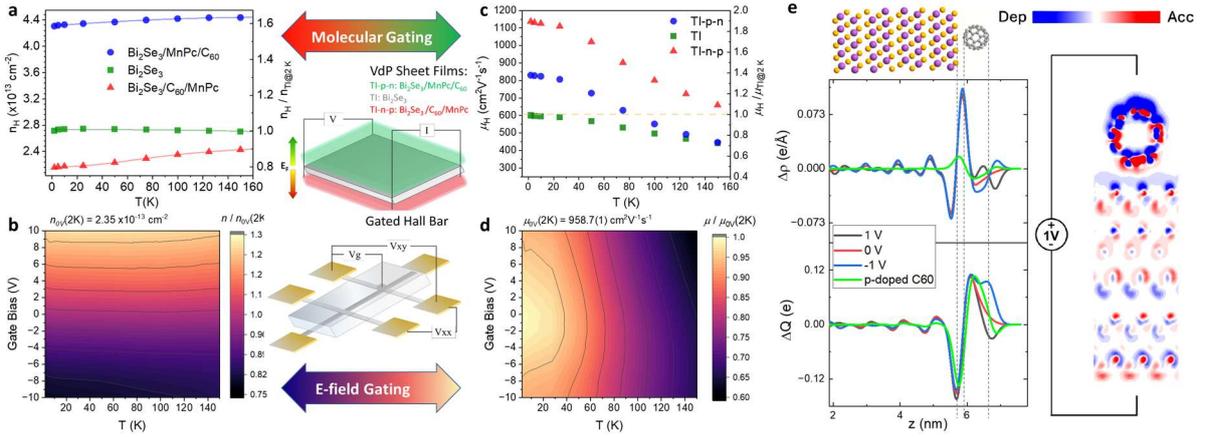

*Figure 2| Molecular gating. Hall effect measurements of a. the effect of n-p and p-n molecular diodes on top of a $Bi_2Se_3$ film 14 nm thick. The n-p bilayer decreases the carrier concentration in the n-type TI, whereas the p-n bilayer increases it. b. A similar effect can be achieved with a top-gate requiring a bias of several volts. c. The carrier mobility is increased by up to almost factor 2 with the molecular bilayers, probably due to the formation of dispersive bands hybridised with the TI surface states. d. The electric bias, independently on the polarity, reduces the carrier mobility at low T. e. DFT modelling of the charge and charge density changes at the $Bi_2Se_3$/$C_{60}$ interface. Top: plane-averaged charge density difference, $\Delta\rho$ (e/Å) vs position. Bottom: integrated charge transfer, $\Delta Q(z) = \int \Delta\rho(z') dz'$. A positive built-in bias of 1 V across the structure or an electron transfer from fullerene to phthalocyanine ($4\times10^{-3}$ holes per C atom) are in agreement with observations of carrier concentration and TEM cross-section images.*

## IV. Weak Anti-localisation and Colossal SOI

Xray Absorption Spectroscopy (XAS) measurements at the $L_{2,3}$ Se edge show changes in the nonmetal electronic structure with the molecular diode. Whereas the signal at the $L_2$ edge is enhanced by the molecules, the $L_3$ region is partly suppressed. The pre $L_3$ edge at 1435 eV has been associated to hole-charged state of Se in ion-doped $Na_xMoSe_2$. [24] This region is enhanced by the $C_{60}$/MnPc (n-p) diode, but not by the MnPc/$C_{60}$ (p-n) bilayer, matching the observed changes in carrier concentration from Hall measurements. Magnetic dichroism has been reported in Mn-doped $Bi_2Se_3$, [25] but we don't observe any dichroism at the Se edge in either diode configuration, probably due to the lack of interdiffusion and weak coupling of the Mn metal ion in phthalocyanine with the TI substrate (Fig. 3b).

Changes in the spin-dependent scattering of $Bi_2Se_3$ cannot therefore be attributed to local moments induced by the molecules in the TI film, but rather to changes to the electronic structure and SOI.

Many papers use the expression derived by Hikami et al (HLN) [26] to interpret the magnetoresistance of TI materials. Although this quantum interference (QI) theory has been shown to be valid for Dirac fermions, [27] there are also important differences highlighted in a recent theory [ref] (LHC) that should be taken into account. For example, in both theories the diffusion coefficient is central, $D = \frac{1}{2} v_F^2 \tau$, where $\tau$ is the lifetime of a momentum eigenstate. In QI this is normally equated to the elastic lifetime but that is only possible where the spin-orbit lifetime is much less than the elastic lifetime and that is not the case here. In the LHC theory, the diffusion coefficient is a function of the strength of the SOI so we have taken the approach of combining both theories and use the LHC expressions to obtain the diffusion coefficient:

$$D = \frac{v_F^2 \tau_o}{2 + 2\lambda + \frac{5}{4}\lambda^2}$$

Where $\lambda$ is proportional to the strength of the SOI and is obtained from fitting the MR and $\tau_o$ is the elastic lifetime. As we are interested in the SOI lifetime, we use the HLN theory to fit the MR as well, but use the LHC definition for D to obtain the lifetimes.

The full HLN expression describing the correction to the Boltzmann conductance in two dimensions where the magnetic field is perpendicular to the plane of the sample is:

$$\Delta g(B) = -\frac{e^2}{2\pi^2 \hbar} \left[ \Psi\left(\frac{1}{2} + \frac{B_1}{B}\right) - \frac{3}{2} \Psi\left(\frac{1}{2} + \frac{B_2}{B}\right) + \frac{1}{2} \Psi\left(\frac{1}{2} + \frac{B_3}{B}\right) \right]$$

Where $\Psi$ is the digamma function and the characteristic fields are:

$$B_1 = B_0 + B_{so} + B_s$$
$$B_2 = \frac{4}{3} B_{so} + \frac{2}{3} B_s + B_i$$
$$B_3 = 2B_s + B_i$$

The suffixes stand for potential or elastic (0), inelastic (i), magnetic (s) and spin-orbit (so) scattering. There are four effects accounted for in the HLN equation, magnetic scattering we have ignored since there is no evidence of magnetic impurities in the samples. The elastic scattering rate was determined from the low temperature resistivity and was not treated as a fitting parameter leaving only the inelastic and spin-orbit scattering times. Figure 3c shows the magnetoresistance (MR) for the three different samples measured at 2 K. The positive MR is a signature of strong spin-orbit scattering and is weak antilocalisation. At 2 K, the inelastic phonon scattering has been largely frozen out, so the inelastic scattering is probably due to electron-electron interactions. The fits return nearly the same value of the inelastic time for all samples which is very reassuring. We can then confidently say that the difference in the MR between the samples is due entirely to the SOI.

A value of $\tau_{so}$ of 0.173 ps is indicative of very strong spin-orbit interaction as expected for $Bi_2Se_3$. We can compare this $\tau_{so}$ with typical values from disordered metals, for example, amorphous $Cu_{50}Ti_{50}$ has a value of 1.0 ps, [28] whereas doping CaAl with Au to increase the SOI produced in $Ca_{70}Al_{28}Au_2$ a value of 0.14 ps. [29,30] The samples with $C_{60}$/MnPc and MnPc/$C_{60}$ have increased the spin-orbit interaction to values of 0.107 ps and 0.085 ps, which is remarkable and comparable to that found in e.g. in $Pd_{80}Si_{20}$ of 0.09 ps. [31] Since the TI material in all samples is the same, we can estimate (using $\hbar\backslash\tau_{so}$) that the addition of a layer of $C_{60}$/MnPc has increased the SOI energy by 2.5 meV and the MnPc/$C_{60}$ layer by 4.2 meV -both with the same sign and despite the opposite change in carrier density.

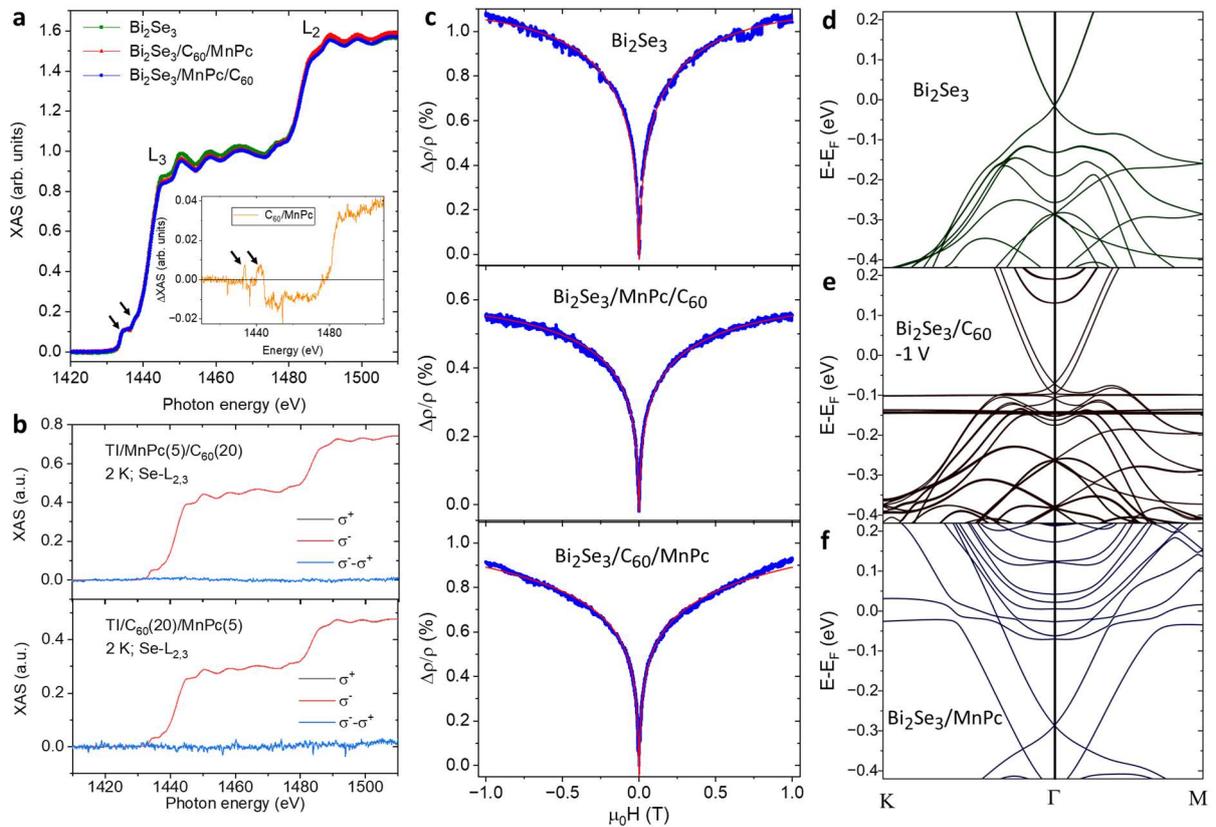

*Figure 3| SOI and electronic structure in molecular gated $Bi_2Se_3$. a. Xray Absorption Spectroscopy (XAS) at Se $L_{2,3}$ edge for $Bi_2Se_3$ pristine and with molecular diodes on top. b. Xray Magnetic Circular Dichroism at the Se L2,3 edge with molecular diodes shows no magnetic order. c. Measurements of weak anti-localisation and fit to the HNL model for $Bi_2Se_3$ pristine (top), with a p-n (middle) and an n-p molecular diode overlayer (bottom). Parameters derived from the fit are in Table 1. DFT simulations of the band structure in d. pristine, e. with a $C_{60}$ interface in which the diode is modelled as a negative voltage and f. for $Bi_2Se_3$/MnPc. The molecular gating leads to the formation of flat bands near the Fermi level, which can result in strongly correlated effects, enhanced SOI and Berry curvature.*

Figs. 3d-f illustrate the electronic band structure evolution along K - Γ - M high-symmetry points for 5 quintuple layers (5QL) pristine $Bi_2Se_3$, $Bi_2Se_3$(5QL)/$C_{60}$ and $Bi_2Se_3$(5QL)/$C_{60}$ with a molecular bias. The pristine $Bi_2Se_3$ shows a well-defined gapless Dirac cone at the Γ-point, characteristic of topological surface states. The bands exhibit a smooth dispersion, confirming the unperturbed topological nature of the system. Adding a $C_{60}$ interface leads to hybridisation near the Dirac point and the emergence of flat bands near the Fermi level, originating from the molecular orbitals. These bands may be partly localised, contributing to an enhanced Berry curvature, [32,33] as observed in topological systems such as twisted graphene and bilayer Kagome metals. [34–37] Taking into consideration the differences in band gaps (1.2 eV in MnPc vs 2.3 eV in $C_{60}$) and associated shift in the HOMO and LUMO levels at the molecular interface, [38] a 1 V built-in bias makes the flat bands much more pronounced, indicating stronger localisation of electronic states. The shift of these states further from the Dirac cone suggests enhanced charge transfer, as observed experimentally when comparing a TI/$C_{60}$ interface with TI/$C_{60}$/MnPc (Supplementary table), and a greater molecular orbital contribution in the electronic structure that results in the stronger SOI.

*Table 1: Carrier parameters of pristine and molecular-gated $Bi_2Se_3$.*

| Parameter @2K | TI | TI/n-p | TI/p-n |
|---|---|---|---|
| $ρ_0$[μΩcm] | 517 ± 3 | 358 ± 2 | 249 ± 1 |
| $n_H$[$10^{13}$ cm$^{-2}$] | 2.708 ± 0.003 | 2.152 ± 0.002 | 4.298 ± 0.004 |
| μ[cm$^2$V$^{-1}$s$^{-1}$] | 601.0 ± 0.8 | 1136 ± 2 | 829 +- 1 |
| D[cm$^2$/s] | 58.8 ± 0.2 | 102.2 ± 0.1 | 112.2 +-0.1 |
| **$τ_{SO}$[$10^{-12}$ s]** | **0.173 ± 0.002** | **0.107 ± 0.002** | **0.085 ± 0.001** |
| **$l_{SO}$ [nm]** | **32** | **33** | **31** |
| $τ_i$[$10^{-12}$s] | 33.5 ± 0.3 | 32.2 ± 0.3 | 21.5 ± 0.4 |
| $τ_0$[$10^{-12}$s] | 0.044 ± .006 | 0.078 ± 0.001 | 0.057 ± 0.002 |
| m.f.p. [nm] | 32.77 ± .04 | 57.56 ± 0.08 | 52.92 ± 0.08 |
| $Θ_D$[K] | 271.4 ± .2 | 265.1 ± 0.2 | 237.9 ± .2 |

V. Vibrational Modes and Optical Tuning

The spin-orbit interaction results in a slow-down of the three main vibrational modes in $Bi_2Se_3$; the out-of-plane modes $A_{1g}^1$ and $A_{1g}^2$ (Raman shifts of 73 and 175 cm$^{-1}$) and the in-plane mode $E_g^2$ (133 cm$^{-1}$), see Fig. 4a. [39–42] In our measurements, $Bi_2Se_3$ wafers partly covered by molecular diodes and probed with a low energy red laser (633 nm) show a slower oscillation frequency for all three modes in both n-p and p-n covered areas, but the effect is stronger for MnPc/$C_{60}$ (n-p) in the $A_{1g}^2$ mode, see Fig. 4b. A softening of the vibrational modes is consistent with a stronger SOI, [43] and may also be linked with changes in the Debye temperature observed in Bloch-Gruneissen fits of the resistivity vs. temperature, see Table 1 and Supplementary Fig. S1. Significantly, this slowdown is a function of the excitation wavelength, with a blue-shift of the vibrational modes in $Bi_2Se_3$/diode samples w.r.t. to pristine $Bi_2Se_3$ as the photon energy increases. The reduced or even inverted effect at higher photon energies may be due to exciton generation across the HOMO-LUMO energy gaps in $C_{60}$ and phthalocyanine (~2.3 and 1.6 eV), leading to a reduced charge transfer and decoupling of the molecular/TI interface.

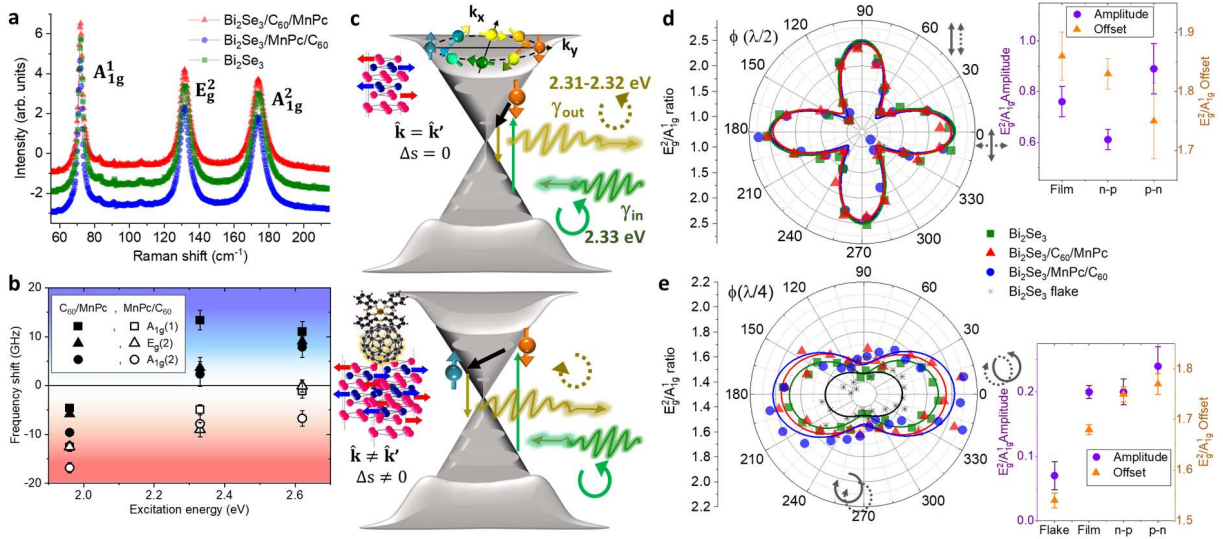

*Figure 4| Raman spectroscopy and optical gating of the SOI. **a.** Raman spectra of $Bi_2Se_3$ with and without molecular bilayers. **b.** The Raman modes are slowed/softened by the molecular diodes. This effect can be cancelled by light exposure above the HOMO-LUMO gap (~2.3 eV). Spectra are averaged over 24 sample points. **c.** Schematic of circularly polarised Raman spectroscopy of the $E_g^2$ (in-plane) vibration. An electron excited from the valence to the conduction band by 532 nm circularly polarised light (2.33 eV) will conserve its spin during the excitation and decay. Some of the energy is transferred to a vibration (~0.02 eV for the 135 $cm^{-1}$ mode). SOI and lattice scattering can flip the electron spin and result in the emission of a photon with opposite polarisation. **d.** Comparison of the $E_g$ to $A_g^2$ intensities with incident linearly polarised light rotated from parallel (Y) to perpendicular (X) to the detector (Y) with a $\lambda/2$ waveplate ($\phi$ polariser angle). **e.** Ratio of in-plane ($E_g^2$) to out-of-plane ($E_g^2$) Raman mode scattering intensity with a left-handed circularly polarised detector as the incident polarisation is changed between right-handed and left-handed with a $\lambda/4$ plate. Lines in e and f are fits to a cos ($2\phi$ and $\phi$) function. Insets in d. and e. are the fit parameters.*

Polarised Raman spectroscopy may also offer a means to probe the spin-orbit interaction and surface states in $Bi_2Se_3$. Due to crystal symmetry, the $A_{1g}$ modes are quenched when measuring with cross-polarised linear light (e.g. excitation in X and probing in Y), and maximised when the polarisation of the excitation and probe are parallel (XX or YY). [44,45] This is the case in all our samples, independently of the coverage of the TI film (Fig. 4d). Molecular gating therefore has no influence on the symmetry of in-plane and out-of-plane vibrations as probed by linearly polarised light. The situation changes when probing the system with circularly polarized light (Fig. 4e). Also for symmetry considerations, measuring in a cross-polarisation setup (e.g. right-handed circular excitation and left-handed circular probing or RL) leads to maximum intensity in the out-of-plane mode $E_g$ and minimum intensity for the in-plane modes ($A_{2g}^1, A_{2g}^2$). [46] When the polarisation handedness is conserved between excitation and probing, e.g. RR, the opposite situation takes place, with a minimum out of plane $E_g$ mode intensity. The situation can be enhanced by spin-orbit coupling during resonant Raman scattering, i.e. when electrons may be promoted from the valence to the conduction band. Due to spin conservation during optical absorption ($\Delta S = 0$), excitation with e.g. left-handed circular polarised light will promote spin-up electrons to a higher energy state. During Raman scattering, some of the electron momentum is transferred to a vibrational mode and the electron will lose energy before decaying back to the lower energy state. Due to spin-momentum locking at the surface of the TI, the emitted photon will only have the same polarisation as the absorbed one only if the electron conserves its momentum direction ($\hat{\mathbf{k}} = \hat{\mathbf{k}}'$). A change of the momentum direction in the XY plane ($\hat{\mathbf{k}} \neq \hat{\mathbf{k}}'$) must be accompanied by spin transfer to the lattice via the SOI and the emission of a photon with a different polarisation. This can happen if the SOI is strong and $\tau_{so}$ is of the same order as the lifetime of the virtual state, typically ~$10^{-13}$-$10^{-14}$ s. [47,48] The cross-polarised emission at the $E_g$ peak will therefore be larger for thin film Bi$_2$Se$_3$ compared to quasi-bulk crystals (flakes). In Bi$_2$Se$_3$ film regions covered by a molecular diode, we can observe that the very short spin-orbit time results in further enhancement of this cross-polarised in-plane mode emission in the RL (or LR) configuration.

## Conclusions

Molecular diodes offer an alternative to gating the electronic states of TI thin films and 2-dimensional materials. This 'gentle' method leads to a tuning in the carrier density and enhanced mobility. Surprisingly, it also results in a strong enhancement of what was already a very strong SOI, with scattering fields and spin-orbit magnetic fields that are significantly larger than reported values. The strengthened spin-orbit interaction can lead to new physics, with tight Berry curvatures and polarised vibrational coupling. This dependence on excitation energy allows for the optical gating of the vibrational frequency.

## METHODS

Deposition: TI-Molecule hybrid thin films were fabricated using a multi-chamber deposition system allowing ultra-high vacuum ($10^{-10}$ mbar) transfer of wafers between individual molecular beam epitaxy reactors. [0001] oriented Al$_2$O$_3$ wafers with dimensions 20 mm x 20 mm were precleaned in acetone and isopropanol alcohol before being outgassed at 500 °C in the UHV chamber before growth. The Bi$_2$Se$_3$ layer was co-evaporated with a Bi:Se flux ratio of 1:40 while maintaining the substrate temperature at 235 °C. After growing 14 nm at 0.15 nm/min the epitaxial material is cooled to 150 °C in a selenium only flux to limit the creation of Se vacancies. Detailed characterisation and methodology for these TI films can be found in previous works. [49] The TI film is then transferred under UHV to separate chamber where the single wafer is masked to create the three unique films. The commercially available MnPc and $C_{60}$ powders were evaporated from low-temperature Knudsen cells at 430 and 500 °C respectively. During the molecule evaporation the substrate was maintained at an ambient temperature, once again, to limit the risk of Se vacancy generation.

Transport: Transport measurements were taken on separated diced 4mm x 4mm wafers of the plain and molecular coated TI films using a Van der Pauw geometry for characterising the Hall coefficients and sheet resistance as a function of measurement temperature. Such organic films are incompatible with traditional fabrication processed to their high solubility. Separate plain films grown with nominally the same process were processed into Hall bars with a 24 nm Alumina gate dielectric grown by ALD to compare the effect of purely E-field gating upon the magneto-transport.

NEXAFS/XMCD: Measurements at the BOREAS beamline of the ALBA Synchrotron.

Raman: Raman spectroscopy was carried out with a Horiba LabRam HR800 instrument using blue (473 nm), green (532 nm) and red (633 nm) laser excitation, and with an 1800 ln/mm grating. Linearly polarised measurements were carried out with red and blue excitation, and circularly polarised measurements with green. Linear polarisers, lambda quarter and lamda half plates were used in the measurements to rotate the excitation and probing polarisations. Peak positions were obtained with red light excitation measuring in 32 different sample points and averaging the final result.

DFT: First-principles calculations were conducted using density functional theory (DFT) as implemented in QuantumATK. [50] The calculations employed a linear combination of numerical atomic orbitals (LCAO) basis set within the generalized gradient approximation (GGA) framework, using norm-conserving pseudopotentials sourced from PseudoDojo. [51] To accurately capture the Dirac cone surface states in the 2D energy-momentum relationship, spin-orbit coupling (SOC) was incorporated via fully relativistic pseudopotentials. Brillouin-zone integrations were performed using a k-point grid generated via the Monkhorst-Pack scheme, [52] with a density of approximately 10 k-points per Å. A plane-wave energy cutoff of 125 Ha was used for the discretised grid, and all structural relaxations were carried out until the maximum force on any atom was below 0.02 eV/Å. To account for inter- and intra-molecular noncovalent interactions, van der Waals (vdW) corrections were applied to the GGA functional. [53] The slab model in the supercell was designed to be infinite and periodic in the x- and y-directions (parallel to the slab surface) but finite along the z-direction (normal to the slab). To eliminate spurious interactions between neighbouring periodic images, a vacuum region exceeding 20 Å was added along the z-direction. In slab calculations with asymmetric surfaces, periodic boundary conditions can introduce an artificial macroscopic electrostatic field. [54] To mitigate this, mixed

Neumann and Dirichlet boundary conditions were applied at the $C_{60}$ and $Bi_2Se_3$ sides of the slab, respectively, serving as an alternative approach for dipole correction. [55] The absence of an artificial electrostatic field at zero applied field was verified by analysing the planar-averaged Hartree potential, where a flat vacuum potential profile confirmed the effectiveness of the correction.

TEM: Electron transparent lamella of the samples were prepared by focused ion beam (FIB) lift-out techniques using a dual beam electron-beam/FIB instrument, a Thermo Fisher Helios Xe Plasma FIB. A 30 kV xenon beam was used to mill into the bulk with currents 6.7 nA and 1.8 nA to extract a section which was subsequently thinned using a current of 74 pA and polished with a 5 kV 47pA ion beam to a minimum thickness of between 55 - 75 nm. STEM and EFTEM imaging was performed in a JEOL ARM200cF at 80 kV.

ACKNOWLEDGEMENTS

We thank the Engineering and Physical Sciences Research Council UK for support via grants EP/S030263/1 and EP/X027074/1. OC and HV acknowledge the support of the EC project INTERFAST (H2020-FET-OPEN-965046). We also acknowledge the support of the Henry Royce Institute for Advanced Materials for access to the Royce Deposition System facilities at the University of Leeds; EPSRC Grant Number EP/P022464/1. T.M. thanks the Royal Academy of Engineering for support via the fellowship RF\201920\19\245.